\documentstyle[twoside,12pt]{article} % had been art12
\font\myaddressfont=cmti10
\input amssym.def
\input amssym.tex
\def\refp#1{(\ref{#1})}
\def\hx{{\hat X}}
\def\dq{{\dot Q}}
\def\lrp#1{\left( #1\right)}
 \newif\ifMarginNotes \MarginNotestrue
\def\mrgn#1{\ifMarginNotes\setbox0=\vtop{\hsize 6.75pc
   {\noindent\relax #1\par}}\leavevmode
   \vadjust{\dimen0=\dp0 \dimen1=\ht0\advance\dimen1 by .5ex
 \advance\dimen0 by -.5ex
  \kern-\dimen1\hbox{\kern\hsize\kern.5pc$\leftarrow$
  \box0}\kern-\dimen0}\fi}
\catcode`\@=11
\font\twelvemsb=msbm10 scaled 1200
\font\tenmsb=msbm10
\font\ninemsb=msbm7 scaled 1200%msbm9
\newfam\msbfam
\textfont\msbfam=\twelvemsb  \scriptfont\msbfam=\tenmsb
  \scriptscriptfont\msbfam=\ninemsb
\def\msb@{\hexnumber@\msbfam}
\def\Bbb{\relax\ifmmode\let\next\Bbb@\else
 \def\next{\errmessage{Use \string\Bbb\space only in math
mode}}\fi\next}
\def\Bbb@#1{{\Bbb@@{#1}}}
\def\Bbb@@#1{\fam\msbfam#1}
\font\bigeufm=eufm10 at 20pt
\font\twelveeufm=eufm10 scaled 1200
\font\teneufm=eufm10

\font\seveneufm=eufm7

\newfam\eufmfam
\textfont\eufmfam=\twelveeufm
\scriptfont\eufmfam=\teneufm
\scriptscriptfont\eufmfam=\seveneufm
\def\frak{\relax\ifmmode\let\next\frak@\else
 \def\next{\errmessage{Use \string\frak\space only in math
mode}}\fi\next}
\def\frak@#1{{\frak@@{#1}}}
\def\frak@@#1{\fam\eufmfam#1}

\catcode`\@=12

\def\ritem#1{\item[{\rm #1}]}

\def\mbni#1{\vskip18truept\noindent{\bf #1}}

\def\la{\lambda}
\def\Ga{\Gamma}
\def\part{\partial}

\def\Mat{{\rm Mat}}

\def\Tr{{\rm Tr}\,}
\def\llangle{\left\langle}
\def\jlo{{\rm JLO}}
\def\rrangle{\right\rangle}
\def\lra#1{\llangle #1\rrangle}
\def\A{{\frak A}}

\def\fj{{\frak J}}
\def\hh{\hat{\cal H}}
\def\hensp#1{\enspace \hbox{#1}\enspace}
\def\ga{\gamma}
\def\la{\lambda}

\def\C{{\Bbb C}}
\def\D{{\cal D}}

\def\CH{{\cal H}}

\def\z{{\frak Z}}
\def\fz{{\frak Z}}
\def\Z{\Bbb Z}
\def\clips{,\ldots}
\def\be{\begin{equation}}
\def\ee#1{\label{#1}\end{equation}}
\def\beq{\begin{eqnarray}}
\def\nn{\nonumber}
\def\eeq{\end{eqnarray}}
\def\l{\left}
\def\r{\right}
\def\lll#1{\lra{\lra{\l\{ #1\r\} }}}

\title{Quantum Invariants\thanks{Work supported
in part by the Department of Energy under Grant DE-FG02-94ER-25228 and
by the National Science Foundation under Grant DMS-94-24344.}}
\author{Arthur Jaffe\\
Harvard University\\
Cambridge, MA 02138, USA}
\date{January 1998}
\begin{document}
\hsize=7truein
\hoffset=-.75truein 
\maketitle
\thispagestyle{empty}
\begin{abstract}
Consider the partition function 
$\z^Q(a,g)={1\over \sqrt\pi}\int^\infty_{-\infty}\Tr_{\CH} 
(\gamma U(g)a e^{-Q^2 + itda}) e^{-t^2}dt$.
In this paper we give an elementary proof that this is an invariant.
This is what we mean: assume that $Q$ is a self-adjoint operator acting on a 
Hilbert space $\CH$, and that the operator $Q$ is odd with 
respect to a \hbox{$\Z_2$-grading} $\gamma$ of $\CH$.  
Assume that $a$ is an operator that is even with respect to $\ga$ and whose 
square equals I. Suppose further that $e^{-Q^2}$ has a finite trace, and that 
$U(g)$ is a unitary group representation that commutes with $\gamma$, with 
$Q$, and with  $a$. Define the differential $da=[Q,a]$.  
Then $\z^Q(a,g)$ is an invariant in the following sense: if 
the operator $Q(\la)$ depends differentiably on a parameter $\la$, 
and if $d_\la a$ satisfies a suitable bound, 
(we specify these regularity conditions completely in \S XI) 
then  $\fz^{Q(\la)}(a,g)$ is independent of $\la$.
Once we have set up the proper framework, a short 
calculation in \S IX shows that $\part\fz^{Q(\la)}/\part\la=0$. 
These considerations apply to non-commutative geometry, to super-symmetric 
quantum theory, to string theory, and to generalizations of these theories 
to underlying quantum spaces.
\end{abstract}

\section{Introduction}
In an earlier paper [QHA] we have studied a class of geometric invariants
that arise within the framework of differential geometry and its
non-commutative generalization [C1, C2, JLO].  By pairing a cocycle $\tau$
with an operator-valued, even, square-root of unity $a$, we obtained a 
specific formula for an invariant $\z^Q(a,g)=\lra{\tau^\jlo,a}$. 
In case that $\tau$ is the JLO-cochain [JLO], this invariant has the 
numerical value 
${1\over \sqrt\pi}\int^\infty_{-\infty}\Tr_{\CH} 
(\gamma U(g)a e^{-Q^2 + itda}) e^{-t^2}dt$, see [QHA].
Here $Q$ is a self adjoint operator and Hilbert space $\CH$, and
$da=[Q,a]$.  
We assume that the differential $Q$ is odd with respect to 
the $\Z_2$-grading $\ga$, while $a$ is even.  We also assume that $g$ is an 
element of a group $G$ of symmetries of $Q$ and of $a$.  
The invariant $\fz$ is not necessarily integer, but it is 
an integer-valued when $g$ equals the identity element.

In this note we present an alternative point of view to [QHA].  Rather than
relating $\fz$ to a general theory of invariants, to entire cyclic
cohomology, and to $K$-theory, we start from the formula for 
$\z^{Q(\la)}(a,g)$ above and ask 
the basic question: can one see directly, when $Q=Q(\la)$ depends on a 
parameter $\la$, that $\z^{Q(\la)} (a,g)$ is actually independent of $\la$?

We answer this question affirmatively, by studying a new auxiliary Hilbert
space $\hh$ containing $\CH$, and defining a new representation for $\fz$ 
as an expectation $\fj(\la,a,g)$ on
$\hh$.  We replace the operator $Q(\la)$ on $\CH$ with the
operator  $q(\la,a)=Q(\la)+\eta a$ on $\hh$.  We call $q$ 
the {\it extended supercharge}.  Here $\hh$ differs from $\CH$ by also 
containing the additional independent fermionic coordinate $\eta$ chosen so 
that $\eta^2=I$ and $\eta Q(\la)+Q(\la)\eta=0$.  We may interpret $\eta a$ 
as a connection associated with the translation in the auxiliary direction 
$t$, paired with $\eta$.  In \S VIII we define an expectation on $\hh$, 
namely
\be
\fj(\la,a,g)= \lra{\lra{Ja}}\;,  
\ee{1.1}
where the notation is explained in (VIII.1--2).  We also show that 
$\fj(\la,a,g)$ and $\fz^{Q(\la)}(a,g)$ agree. 

Once we have defined the proper framework, we show with
a short calculation in \S IX that $\fz$ is constant. We 
present the algebraic aspects of the proof (because these are new)     
without any analytic details.  Of course the analytic details are
absolutely crucial.  Taking 
$\la\in [0,1]$, we could set $Q(\la)=\la Q+(1-\la)\tilde Q$ 
and interpolate between any $Q$ and any other $\tilde Q$, as long as they
both commute with the symmetry group $U(g)$.  
Thus were algebraic arguments presented here valid without any 
further assumptions, we would be in the unfortunate position of 
showing that all invariants for a given $a$ agree! 

Though the analytic part of the argument is crucial,
it also remains identical to our paper [QHA].  In that other work, we 
formulate precisely two regularity conditions: first the 
regularity of $Q(\la)$ with respect to $\la$, and secondly the regularity
of $a$ with respect to $Q(\la)$.  We call the latter the
fractional differentiability properties of $a$. 
The conditions we give are useful because they are easy to verify in a 
large set of examples.  

Under these regularity conditions, 
$\fz^{Q(\la)}(a,g)$ is once-differentiable in $\la$.  Furthermore, 
the resulting $\la$-derivative of $\fz$ equals 
the expression that we would obtain by interchanging the order of
differentiating and the order of 
taking traces or integrals in the definition of $\fz$.
The analysis to establish these facts is lengthy, but can be taken over
directly from [QHA].   For the convenience of the reader, in \S XI, we 
summarize the analytic hypotheses used in argument.

In \S X we consider a different but related case with two differentials 
$Q_1$ and $Q_2$, but where only $Q_1$ is invariant under the symmetry 
group $G$. Instead we assume that $Q^2_1 +Q^2_2$ is also invariant, and that
$Q^2_1-Q_2^2$ commutes with all relevant operators.  
We show in this case that an expectation \refp{10.5} has a representation 
similar to \refp{1.1} and also is an invariant with respect to $\la$.

\section{The Supercharge}
\setcounter{equation}{0}
Our basic framework involves an odd, self adjoint operator $Q$ on a
$\Z_2$-graded Hilbert space $\CH$.  This means we have a self-adjoint
operator $\ga$ on $\CH$ for which $\ga^2 = I$.  Thus $\CH$ splits into the
direct sum $\CH=\CH_+ \oplus \CH_-$ of eigenspaces of $\ga$.  The
statement that $Q$ is odd means $Q\ga + \ga Q=0$.  In terms of the direct
sum decomposition,
\be
Q=\pmatrix{0 & Q^\ast_+ \cr
Q_+ &0} \hensp{and} \gamma =\pmatrix{I&0\cr 0&-I} \; . \ee{2.1}
The operator $Q$ is the supercharge\footnote{We are not concerned with
the basic structure of $\CH$ or $Q$, aside from the possibility to perform
the construction in \S V.} and its square
\be
H=Q^2 = \pmatrix{ Q^\ast_+ Q_+ &0\cr 0&Q_+ Q^\ast_+} \ee{2.2}
will be referred to as the Hamiltonian.  We let $x^\gamma=\gamma x
\ga$ denote the action of $\ga$ on operators.  We say that the operator $x$
is even (bosonic) if $x=x^\ga$ and odd (fermionic) if $x^\ga=-x$.  We define
the graded differential
\be
dx = Qx-x^\ga Q\; . \ee{2.3}

We suppose that there is a compact Lie group $G$ with a continuous unitary
representation $U(g)$ on $\CH$ such that 
\be
U(g)\ga = \ga U(g)\;, \qquad \hensp{and}\qquad U(g)Q=QU(g)\;.
\ee{2.4} 
Denote the action of $U(g)$ on the operator $x$ by 
\be
x \to x^g = U(g) x U(g)^{-1} 
\ee{2.5}

\section{The Observables}
\setcounter{equation}{0}
We also consider an algebra of bounded operators $\A$ on $\CH$ with the
properties that each $a\in\A$ is even and invariant.  In other words, each
$a\in\A$ commutes with $\gamma$ and with $U(g)$ for all $g\in G$.
We also consider $\Mat_n(\A)$, the set of $n\times n$ matrices with matrix 
elements (or entries) in $\A$.  If $a,b\in\Mat_n(\A)$ are matrices with 
entries $a_{ij}, b_{ij} \in \A$, we use the shorthand $ab$ to denote the 
matrix with entries $\sum_{k=1}^n a_{ik}b_{kj}\in\A$. 

The differential of elements of $\A$ is 
\be
da = Qa-aQ=[Q,a] \ee{3.1}
is always defined as a quadratic form on $\CH$.  
We make precise the nature of the differentiability of $\A$, in \S XI. 
The operators $\A$ and the derivative $d$ are the
fundamental building blocks of non-commutative geometry; here $\A$
generalizes the notion of functions on a manifold $M$, and $Q$ generalizes a
Dirac operator on a bundle over $M$.

\section{The Invariant  {\bigeufm Z}$^{Q(\la)}(a,g)$}
\setcounter{equation}{0}
In [QHA] we gave a simple formula for an invariant.  Let $Q(\la)$ depend on a
real parameter $\la$.  We denote the graded commutator \refp{2.3} 
of $Q(\la)$ with $x$ by  
\be
d_\la x= Q(\la)x-x^\ga Q(\la)\;.
\ee{4.0}
For $a\in \A$ define
\be
\fz^{Q(\la)}(a,g) = \frac1{\sqrt\pi} \int^\infty_{-\infty} \Tr_{\CH} \lrp{\ga
U(g)ae^{-Q(\la)^2 + itd_\la a}} e^{-t^2}dt\;.\ee{4.1}
More generally, we let $a\in \Mat_n(\A)$.  In this case
\be
\fz^{Q(\la)}(a,g) = \frac1{\sqrt\pi} \int^\infty_{-\infty} \Tr_{\CH\otimes
\C^{n^2}} \lrp{\ga U(g)a e^{-Q(\la)^2 + itd_\la a}}e^{-t^2}dt\; , \ee{4.2}
where we extend $\ga,U(g),Q(\la)^2$ to diagonal
$n\times n$ matrices.

\mbni{Theorem I.} {\it For $a\in \A$, assume $a^2=I$. 
Furthermore assume that $Q=Q(\la)$ and $d_\la a=[Q(\la),a]$ satisfy the
regularity hypotheses given in \S XI.  Then $\fz^{Q(\la)}(a,g)$ is 
independent of $\la$.
}
\medbreak
The main point of this paper is to present a new, elementary proof
of Theorem I.

\section{The Extended Supercharge $q$}
\setcounter{equation}{0}
In order to exhibit our proof, we introduce a new Hilbert space $\hh$ on
which the operators $Q,\ga,\A$, and $U(g)$ also act.  In addition, on $\hh$
there are two additional self adjoint operators $\eta$ and $J$, both of
which have square one,
\be
\eta^2 = J^2=I\;, \ee{5.1}
and such that
\be
[\eta,x] = [J,x] = 0 \hensp{for} x=\ga, a\in \A,\hensp{or} U(g)\;,\ee{5.2}
and also
\be
\eta J+J\eta = \eta Q+Q\eta = [J,Q] = 0\; . \ee{5.3}
Let $\Gamma = \ga J$ denote a
$\Z_2$-grading on $\hh$, and for $x$ acting on $\hh$ let
\be
x^\Ga = \Ga x \Ga\; . \ee{5.5}
The operator $\eta$ is our auxiliary fermionic coordinate, and
$J=(-I)^{N_\eta}$ is the corresponding $\Z_2$ grading.
\footnote{Suppose that
$\CH= \CH_b\otimes \CH_f$ is a tensor product of bosonic and fermionic
Fock spaces, that $Q$ is linear in fermionic creation or annihilation
operators, and that $\ga= (-I)^{N_f}$.  This would be standard in the physics
of supersymmetry.  Suppose in addition that $\eta=b +b^\ast$ denotes one
fermionic degree of freedom independent of those in $\CH_f$ and acting on
the two-dimensional space $\CH_\eta$.  Then take $\hh=\CH_b\otimes
(\CH_f\wedge \CH_\eta)$ and $J=(-I)^{b^\ast b}$, with $Q,\ga,a$, and $U(g)$
acting on $\hh$ in the natural way.  This gives a realization of (V.1--3) on
$\hh$.} 

Given $a\in \A$,
define the extended supercharge $q=q(\la,a)$ by
\be
q = q(\la,a) = Q(\la) +\eta a\;, 
\ee{5.4}
and also let 
\be
h=h(\la,a)=q(\la,a)^2 = Q(\la)^2 +a^2-\eta d_\la a\;.
\ee{5.4b}
Note that
\be
q^\Ga = -q\;,\hensp{and} h^\Ga=h\; . 
\ee{5.6}
We use the notation $d_q$ to denote the $\Ga$-graded commutator on
$\hh$,
\be
d_q x=qx-x^\Ga q\; . \ee{5.7}
If we need to emphasize the dependence of $q$ on $\la$ or $a$, then 
we write $d_{q(\la, a)} x$.
We continue to reserve $d$ or $d_\la$ to denote the $\ga$-graded commutator
\refp{4.0}.  

\section{Heat Kernel Regularization on $\hh$}
\setcounter{equation}{0}
Let us introduce the heat kernel regularizations $\hat X_n$ of $X_n$ on
$\hh$.  Let $X_n = \{x_0\clips, x_n\}$ denote an ordered set of $(n+1)$
linear operators $x_j$ acting on $\hh$.  We call the $x_j$ {\it vertices} and
$X_n$ a set of vertices.  Choose $a\in \A$ and let 
$q(\la,a)=Q(\la)+\eta a$, and $h=h(\la,a)=q(\la,a)^2$.  
Define the heat kernel regularization $\hat X_n(\la,a)$ of 
$X_n=\{x_0\clips,x_n\}^\wedge (\la, a) $ by
\be
\hx_n(\la,a) = \int_{s_j>0} x_0e^{-s_0
h}x_1 e^{-s_1h}\cdots x_n e^{-s_n h} \delta (1-s_0 - \cdots - s_n) ds_0
\cdots ds_n\; . \ee{6.1}
Note that if $T$ is any operator on $\hh$ that commutes with $h=q^2$, then
\be
\{ x_0 \clips, x_j T,x_{j+1} \clips , x_n\}^\wedge (\la,a) = \{
x_0\clips,x_j,Tx_{j+1}\clips,x_n\}^\wedge (\la,a)\; . \ee{6.2}
Furthermore $T=J\eta$ anti-commutes with $q(\la,a)$ and
commutes with $h(\la,a)$ for all $a$.

\mbni{Proposition II.} (Vertex Insertion) 
{\it Let $X_n = \{x_0\clips,x_n\}$ denote a set of
vertices possibly depending on $\la$.  Then with the notation
$\dq = \part Q(\la)/\part \la$, we have 
\beq
{\part\over \part\la} \{ x_0 \clips ,x_n\}^\wedge (\la, a) &=& -\sum^n_{j=0}
\{x_0\clips, x_j,d_q \dq , x_{j+1}\clips, x_n\}^\wedge (\la, a) \nn\\
&&+ \sum^n_{j=0} \{x_0\clips, {\part x_j\over \part\la}\clips,
x_n\}^\wedge (\la, a) \; . \label{6.3}
\eeq
Here
\be
d_q \dq = d_{q(\la,a)} \dq = d_\la \dq + \eta[a,\dq]\; . \ee{6.4}
}
\mbni{Proof.} By differentiating $\hx_n$ defined in \refp{6.1}, we obtain
two types of terms.  Differentiating the $x_j$'s gives the second sum in
\refp{6.3}.  (This sum is absent if the $x_j$'s are $\la$-independent.)  The
other terms arise from differentiating the heat kernels.  We use the identity
\be
{\part\over \part \la} e^{-sh} = -\int^s_0 e^{-uh} {\part h\over \part\la}
e^{-(s-u)h} du \; . \ee{6.5}
Note that
$$
{\part h\over \part \la} = {\part \over \part \la} (q^2) = q {\part q\over
\part \la}  + {\part q\over \part \la} q=d_q \lrp{\part q\over \part \la} =
d_q \dq\; . 
$$
Explicitly
$$
d_q \dq = (Q+\eta a) \dq + \dq (Q+\eta a) = d_\la \dq + \eta [a\dq]\; . $$
Inserted back into the definition of $\hx_n$, we observe that
the differentiation of the heat kernel between vertex $j$ and vertex $j+1$
produces one new $-d_q \dq$ vertex at position $j+1$.  This completes the
proof of (VI.3).
\medbreak
Define the action of the grading $\Ga$ on sets of vertices $X_n$ by
\be
X_n \to X^\Ga_n = \{ x^\Ga_0,x^\Ga_1\clips, x^\Ga_n\}\; . \ee{6.6}
Since $q^2 = (q^\Ga)^2$, the regularization $X_n\to \hx_n$ commutes with
the action of $\Ga$, namely
\be
\lrp{\hx_n(\la, a)}^\Ga = \lrp{X^\Ga_n}^\wedge (\la, a)\; . \ee{6.7}
It is also convenient to write explicitly differential of $\hx_n$,
\be
d_q \hx_n(\la, a) = q\hx_n(\la, a) - \hx_n(\la, a)^\Ga q 
= \{ q x_0 ,x_1\clips,
x_n\}^\wedge (\la, a) - \{ x^\Ga_0\clips,x^\Ga_n q\}^\wedge (\la, a) \;. 
\ee{6.8}
Note that
\be
d_q \hx_n (\la, a) = \sum^n_{j=0} 
\{ x^\Ga_0,x^\Ga_1\clips,x^\Ga_{j-1} , d_q x_j \clips,x_n\}^\wedge(\la, a)\;. 
\ee{6.9}
One other identity we mention is
\mbni{Proposition III.} (Combination Identity) {\it The heat 
kernel regularizations satisfy 
\be
\{ x_0 , x_1\clips, x_n\}^\wedge (\la, a) 
= \sum^n_{j=0} \{x_0,x_1\clips,x_j , I,x_{j+1}\clips, x_n\}^\wedge (\la, a)\; . 
\ee{6.10}
}
\mbni{Proof.} The $j$th term on the right side of \refp{6.9} is
\beq
&& \hskip-.75in
\{x_0\clips,x_j ,I, x_{j+1}\clips, x_n\}^\wedge (\la, a)\nn\\
&& = \int_{s_j>0} x_0
e^{-s_0 h}\cdots x_je^{-(s_j+s_{j+1})h}\cdots x_n e^{-s_{n+1}h}
\delta(1-s_0-\cdots -s_{n+1}) ds_0\cdots ds_{n+1}\; . \label{6.11} 
\eeq
Change the $s$-integration variables to $s'_0 = s_0,s'_1=s_1 \clips,
s'_j=s_j+s_{j+1}$, $s'_{j+1}=s_{j+2}\clips,s'_n=s_{n+1}$, and
$s'_{n+1}=s_j$.  This change has Jacobian 1, and the resulting integrand has
the form of the integrand for $\{x_0\clips,x_n\}^\wedge$ with variables
$s'_0\clips,s'_n$, namely
\be
\int_{s'_0, s'_1, \ldots, s'_n >0} ds'_0\cdots ds'_{n}
 \l(\int ds'_{n+1}
x_0 e^{-s'_0 h}\cdots  x_n e^{-s'_{n}h}   
\delta(1-s'_0-\cdots -s'_{n})\r) \;,
\ee{6.11b}
with the integrand depending on the variable $s'_{n+1}$ only through 
the restriction of the range of the $s'_{n+1}$ integral. 
The original domain of integration restricts $s'_{n+1}$ to the range 
$0\le s'_{n+1}\le s'_j$, so the dependence of the integrand on $s'_{n+1}$
is the characteristic function of the interval $[0,s'_j]$.  
Thus performing the $s'_{n+1}$ integration 
produces a factor $s'_j$ in the $s'_0, \ldots, s'_n$-integrand.  
Add the similar results for $0\le j\le n$ to give the factor 
$s'_0 +s'_1 + \cdots s'_n$. But the delta function in \refp{6.11b} restricts
this sum to be 1, so the integral of the sum is 
exactly $\{x_0\clips, x_n\}^\wedge(\la,a)$.

\section{Expectations on $\hh$}  
\setcounter{equation}{0}
Let $a\in \A$ satisfy $a^2=I$, and let $\hx_n = \hx_n(\la, a)$ denote the 
heat kernel regularization of $X_n$.  We define the expectation
\be
\lra{\lra{\hx_n}}_{\la,a,g} = {1\over \sqrt{4\pi}} \int^\infty_{-\infty}
\Tr_{\hh} \lrp{\Ga U(g) \hx_n (\la, ta)} dt\; . 
\ee{7.1}
Here we choose $a^2=I$ to ensure that the $t^2$ term in $h$ provides a
gaussian convergence factor to the $t$-integral.  This integral represents
averaging over $a$'s whose squares are multiples of the identity.

These expectations can be considered as $(n+1)$-multilinear expectations
on sets $X_n$ of vertices.  We sometimes suppress the $\la$- or $a$- or
$g$-dependence of the expectations, or the $n$-dependence of sets of
vertices.  Furthermore, where confusion does not occur we omit the
$^\wedge$ that we use to distinguish a set of vertices $X$ from the heat
kernel regularization of the set.
Thus at various times we denotes $\lra{\lra{\hx_n}}_{a,g}$ by
$\lra{\lra{X}}$, or when we wish to clarify the dependence on $n,a$, or $g$
with some subset of these indices, or even as one of the following:
\be
\lra{\lra{\hx_n}}_{\la,a,g} = \lra{\lra{X}}= \lra{\lra{X}}_n =
\lra{\lra{X}}_{n,a}=\lra{\lra{X}}_{n,a,g}\; , \ee{7.2}
etc.
\mbni{Proposition IV.} {\it With the above notation, we have the identities}
\vglue16truept
\noindent \hbox{($\Ga$-invariance)}
\vglue-30truept
\be  \lra{\lra{X}}_n = \lra{\lra{X^\Ga}}_n\; , \ee{7.3}

\vglue6truept\noindent \hbox{(differential)}  \vglue-34truept
\be
 \lra{\lra{d_qX}}_n = \sum^n_{j=0} \lra{\lra{\l\{
x^\Ga_0,x_1^\Ga\clips,x^\Ga_{j-1},d_qx_j\clips,x_n\r\}}}\; , \ee{7.4}
\vglue6truept \noindent
\hbox{(cyclic symmetry)} \vglue-30truept\be
\lra{\lra{\l\{x_0,x_1\clips,x_n\r\}}}= \lra{\lra{\l\{
x^{g^{-1}\Ga}_n ,x_1,x_2\clips,x_{n-1}\r\}}}\; , \ee{7.5}
\vglue6pt \noindent and
\vglue10truept\noindent \hbox{(combination identity)} 
%\vglue-24truept
\vglue-18truept
\be
\hskip.5truein\lra{\lra{\l\{x_0,x_1\clips,x_n\r\}}}_n =
\sum^n_{j=0} \lra{\lra{\l\{x_0,x_1\clips,x_j,I,x_{j+1}\clips,
       x_n\r\}}}_{n+1}\; . 
\ee{7.6}
Also, in case $Q=Q^g$ and $a=a^g$, then $q=q^g$ and we have 
\vglue16truept\noindent  \hbox{(infinitesimal invariance)} 
\vglue-30truept
\be
  \lra{\lra{d_{q(\la, ta)}X}}=0\; . 
\ee{7.7}
\medbreak
\mbni{Proof.} The symmetry \refp{7.3} is a consequence of the fact that 
$\Ga^2=I$, and $\Ga$ commutes with $U(g)$ and with $q^2$.  The expectation 
of \refp{6.9} completes the proof of \refp{7.4}.  The proof of \refp{7.5} 
involves cyclicity of the trace.  The
identity \refp{7.6} is the expectation of \refp{6.10}.  To establish 
\refp{7.7}, note that
every $\hx_n$ can be decomposed uniquely as 
$\hx_n = \hx_n^{+}+\hx^{-}_n$, where
$\lrp{\hx^\pm_n}^\Ga = \pm \hx^\pm_n$.  The symmetry \refp{7.3} ensures that
$\lra{\lra{d_{q(\la,ta)}X^+_n}}=0$.  On the other hand, $q^\Ga = -q$, 
together with cyclicity of the trace and $q^g = q$ ensures that
\beq
\lra{\lra{d_{q(\la,ta)}X^-_n}} 
&=& \lra{\lra{q(\la,ta)X^-_n}} 
    + \lra{\lra{X^-_n q(\la,ta)}} \nn\\
&=& \lra{\lra{q(\la,ta)X^-_n}}
    + \lra{\lra{q(\la,ta)^{g^{-1}\Ga}X^-_n}}=0\; . \nn
\eeq

Except in \refp{7.7}, we have implicitly assumed that the vertices 
$x_j$ in $X_n$ are $t$-independent. 
In case that $X_n$ has one factor linear in $t$, 
the heat kernel regularizations of the following agree,
\be
\l\{ tx_0,x_1\clips,x_n \r\}^\wedge (\la, ta) 
= \l\{ x_0,x_1\clips,tx_j\clips,x_n\r\}^\wedge (\la, ta)\; ,
\ee{7.8}
for any $j=0,1\clips,n$.  We then obtain an interesting relation for 
expectations,
\mbni{Proposition V.} (Integration by parts) {\it Let $a^2=I$.  Then}
\be
\lra{\lra{\l\{ tx_0,x_1\clips,x_n\r\} }}_n = \sum^n_{j=0} \frac12
\lra{\lra{\l\{x_0\clips,x_j,\eta d_\la a,x_{j+1}\clips,x_n\r\}}}_{n+1}
\; . \ee{7.9}

\mbni{Proof.} In order to establish \refp{7.9}, we collect
together the terms $\exp(-s_jt^2)$ that occur in
$\l\{ x_0\clips,x_n\r\}^\wedge (ta)$.  Since the integrand
for the heat kernel regularization has a $\delta$-function restricting
the variables $s_j$ to satisfy $s_0 + \cdots +s_n=1$, we
obtain the factor $\exp(-t^2)$.  Write
$$
t e^{-t^2} = -\frac12 \frac{d}{dt} \lrp{e^{-t^2}} 
$$
and integrate by parts in $t$.  The resulting derivative involves
the $t$-derivative of each heat kernel $\exp-(s_jq (\la,ta)^2)$
with the quadratic term in $t$ removed from $q^2$.  Note that 
\beq
e^{-st^2} {d\over dt} e^{-s(q^2-t^2)}
&=& -\int^s_0 e^{-uq^2} \lrp{ {d\over dt} (q^2-t^2)}
e^{-(s-u)q^2} du \nn\\
&=& \int^s_0 e^{-uq^2} \eta d_\la a e^{-(s-u)q^2} du \; . \nn\eeq
Here we use \refp{5.4b} with $ta$ replacing $a$ and      
with $a^2=I$ in order to evaluate the $t$ derivative of $q^2-t^2$.
Thus each derivative introduces a new vertex equal to
$\frac12 \eta d_\la a$, and the proof of \refp{7.9} is complete.

\section{The Functional {\bigeufm J}$(\la,a,g)$}
\setcounter{equation}{0}
Let us consider a single vertex and $X_0 = x_0=Ja$, where
$a\in \A$, and its expectation
\be
\fj(\la,a,g) = \lra{\lra{Ja}}\; . \ee{8.1}
Explicitly
\be
\fj (\la,a,g) = {1\over \sqrt{4\pi}} \int^\infty_{-\infty}
\Tr_{\hh} \lrp{\gamma U(g)ae^{-q(\la,ta)^2}} dt\; .
\ee{8.2}
This functional allows us to recover the functional $\fz$.

\mbni{Theorem VI.} {\it Let $a$ satisfy $a^2=I$.  Then}
\be
\fj (\la,a,g) = \fz^{Q(\la)} (a,g)\; . \ee{8.3}
\mbni{Proof.} Let $h=h_0 - t\eta da$, where $h_0 = Q(\la)^2+t^2$.  
The Hille-Phillips perturbation theory for semi-groups can
be written
\beq
&& \hskip -.5in 
e^{-q(\la, ta)^2} = e^{-h_0 +t\eta da}  \nn\\
&& \hskip -.5in \phantom{e^{-q(\la, ta)^2}}
 =  e^{-h_0} +\sum^\infty_{n=1} t^n  \int_{s_j>0}
e^{-s_0h_0}\eta d_\la a e^{-s_1h_0}\eta d_\la a
\cdots \eta d_\la a e^{-s_nh_0}
\delta( 1-s_0-s_1-\cdots - s_n) ds_0ds_1\cdots ds_n\;. \nn\\
&& \label{8.4} \eeq
In the $n$th term we collect all factors of $\eta$ on the left.  
Note that $\eta$ commutes with $a$ and $h_0$, and it anti-commutes with 
$d_\la a$.  Therefore the
result of collecting the factors of $\eta$ on the left is 
$\eta^n (-1)^{n(n-1)/2}$.  If $n$ 
is odd, then $\eta^n=\eta$ and $\Tr_{\CH_\eta}(\eta)=0$.  
Thus only even $n$ terms
contribute to \refp{8.2}.  For even $n$,
$\eta^n(-1)^{n(n-1)/2}=(-1)^{n/2}I$ 
and $\Tr_{\CH_\eta}(I)=2$.  Thus \refp{8.2} becomes 
\be
\fj (\la,a,g) 
= {1\over \sqrt\pi} \int^\infty_{-\infty} dt 
\sum^\infty_{n=0} (-t^2)^n \lra{
\l\{a,d_\la a\clips,d_\la a\r\}}_{2n} e^{-t^2}\; , \ee{8.5}
where we use expectations $\lra{\enspace}_n$ on $\CH$ similar to
$\lra{\lra{\enspace}}_n$ on $\hh$ (but without the $t$-integration) 
and defined by 
%\beq
%\hskip -.25in 
%\lra{\l\{x_0\clips,x_n\r\}}_n &=& \int_{s_j>0} \Tr_{\CH} \lrp{ \gamma 
%U(g)x_0 e^{-s_0 Q(\la)^2}\cdots x_n e^{-s_nQ(\la)^2}} \nn\\
%&& \qquad\qquad\qquad
%\times\delta (1-s_0-s_1-\cdots-s_n) ds_0 ds_1\cdots ds_n \; . 
%\label{8.6}\eeq
\be
\lra{\l\{x_0\clips,x_n\r\}}_n = \int_{s_j>0} \Tr_{\CH} 
\lrp{ \gamma U(g)x_0 e^{-s_0 Q(\la)^2}\cdots x_n e^{-s_nQ(\la)^2}}
\delta (1-s_0-s_1-\cdots-s_n) ds_0 ds_1\cdots ds_n \;. 
\ee{8.6}
But using the Hille-Phillips formula once again, \refp{8.6} is just
\be
\frac1{\sqrt\pi}\int_\infty^\infty dt \Tr_{\CH} \lrp{\gamma U(g) a 
e^{-Q(\la)^2 +it d_\la a }}e^{-t^2} = \fz^{Q(\la)}(a,g)\;.
\ee{8.7}
(Here we use the symmetry of \refp{8.7} under $\ga$  to justify 
vanishing of terms involving odd powers of $d_\la a$.)  Thus we can prove 
that $\fz^{Q(\la)}(a,g)$ is independent of $\la$ by showing 
that $\fj(\la,a,g)$ is constant in $\la$.

\section{ {\bigeufm J}$(\la,a,g)$ Does Not Depend on $\la$}
\setcounter{equation}{0}
We now prove Theorem I.  Calculate $\part \fj/\part\la$ using \refp{6.3}, 
in the simple case of one vertex independent of $\la$.  Thus
\be
{\part\over \part \la} \fj (\la,a,g) 
= {\part\over \part \la}\lra{\lra{Ja}} = - \lra{\lra{
\l\{ Ja,d_{q(\la,ta)}\dq \r\} }}\; . 
\ee{9.1}
Using the identity \refp{7.7} in the form
\be
0 = \lra{\lra{d_{q(\la,ta)} \l\{ Ja,\dq\r\} }} 
= \lra{\lra{ \l\{ d_{q(\la,ta)} (Ja),\dq\r\} }}+\lra{\lra{ \l\{ Ja
, d_{q(\la,ta)} \dq\r\} }}\; , 
\ee{9.2}
we have
\be
{\part\over \part \la} \fj (\la,a,g)
= \lra{\lra{\l\{ d_{q(\la,ta)} (Ja),\dq\r\} }}\; . 
\ee{9.3}
It is at this point that we have used $q^g = q$, namely the invariance 
of both $Q$ and $a$ under $U(g)$.  To evaluate \refp{9.3}, note that
\be
d_{q(\la,ta)} (Ja) = [q(\la, ta),Ja] = Jd_\la a-2t J\eta\; . \ee{9.4}
Here we use the assumption $a^2 = I$.  From Proposition V we therefore infer
\beq
{\part\over \part \la} \fj (\la,a,g) &=& \lra{\lra{\l\{ Jd_\la a,\dq\r\} }} 
- 2 \lra{\lra{\l\{tJ\eta,\dq\r\}}} \nn\\
&=& \lra{\lra{\l\{ Jd_\la a,\dq\r\} }} 
- \lra{\lra{\l\{ J\eta,\eta d_\la a,\dq\r\}}} 
- \lra{\lra{\l\{ J\eta,\dq,\eta d_\la a\r\}}} \; . 
\label{9.5}\eeq

Since $J\eta$ commutes with $h=q^2$, and since $J\eta \dq = -\dq J\eta$, use
\refp{6.2} to establish
\beq
  \lll{J\eta,\eta d_\la a,\dq} + \lll{J\eta,\dq,\eta d_\la a}&  
= &\lll{I,Jd_\la a,\dq} 
- \lll{I,\dq,Jd_\la a} \nn\\ & 
= &\lll{Jd_\la a,\dq,I} + \lll{Jd_\la a,I,\dq}\; .
\label{9.6}
\eeq 
In the last step we also use $\dq^\Ga = - \dq$ and the cyclic symmetry
\refp{7.5}.  Hence we can simplify \refp{9.6} to $\lll{J d_\la a,\dq}$, 
by applying the combination identity
\refp{7.6}.  Substituting this back into \refp{9.5}, we end up with
\be
{\part\over \part \la} \fj (\la,a,g) 
= \lll{J d_\la a,\dq} - \lll{J d_\la a,\dq} = 0\; .
\ee{9.7}
Thus $\fj(\la,a,g)$ is invariant under change of $\la$, and the 
demonstration is complete.

\setcounter{equation}{0}
\section{Independent Supercharges $Q_j(\la)$}
Let us generalize our consideration to the case that there are two 
self-adjoint operators $Q_1 = Q_1(\la)$ and $Q_2 = Q_2(\la)$ on $\CH$ 
such that
\be
Q_1\ga +\ga Q_1 = Q_2\ga + \ga Q_2 = Q_1Q_2 +Q_2Q_1 =0\; . \ee{10.1}
Thus we have two derivatives $d_ja = [Q_j,a]$.  We assume that the 
energy operator on $\CH$ is defined by
\be
H=H(\la) = \frac12 \lrp{Q_1 +Q_2}^2 = \frac12 \lrp{Q^2_1+Q^2_2}  
\ee{10.2}
and that the operator
\be
P = \frac12 \lrp{Q^2_1 - Q^2_2} \ee{10.3}
has the properties:
\begin{itemize}
\item[i)] $P$ does not depend on $\la$.
\item[ii)] $P$ commutes with $Q_1,Q_2$ and with each $a\in \A$.
\item[iii)] $U(g)$ commutes with $Q_1$ and with $H(\la)$.
\end{itemize}
\noindent Assumption (i) corresponds to a common situation where $P$ can be
interpreted as a ``momentum'' operator.  Then the energy, but not the 
momentum is assumed to depend on $\la$.  Assumption (ii) says 
that $Q_1,Q_2$ are translation invariant, and that $\A$ is 
a ``zero-momentum'' or translation-invariant subalgebra. 
According to assumption (iii), $U(g)$ commutes with $Q^2_2$, but $U(g)$ may 
not commute with $Q_2$.  Under these hypotheses, and with appropriate 
regularity assumptions, we showed in [QHA] that for $a=a^g$ and $a^2 =I$,
\be
\fz^{Q_j(\la)} (a,g) 
= {1\over \sqrt\pi} \int^\infty_{-\infty} \Tr \lrp{ \ga
U(g)ae^{-H+itd_1a-t^2}} dt 
\ee{10.5}
is independent of $\la$.  Here we show how the framework above can 
also be used to show that \refp{10.5} is constant.

We introduce on $\hh$ two extended supercharges 
$q_1 = q_1(\la,a) = Q_1+\eta a$ and
$q_2 = Q_2$.  
Thus with $\eta$ as before, 
$\eta Q_1+Q_1 \eta = \eta Q_2+Q_2\eta=0$.  
Define
\be
h=h(\la,ta)=H(\la) + t^2a^2- t\eta d_1 a\; . 
\ee{10.6}
Note that
\be
h=q_1 (\la,ta)^2-P = Q_1(\la)^2 +t^2 a^2 -t\eta d_1 a -P\; , \ee{10.7}
so we can eliminate $Q_2(\la)$ from $h$ by introducing the operator $P$,
that commutes with $a$, $\gamma$, $J$, $U(g)$, $\eta$, and $Q_j(\la)$.

Thus $P$ commutes with all operators that we consider on $\hh$, so we  
repeat the constructions of \S V--IX. However, we replace $q(\la, ta)^2$ in 
the previous construction with $h(\la, ta)$ defined by \refp{10.6}.
Also we replace $d_q x$ with $d_{q_j} x = q_j x -x^\Ga q_j$.
We use the heat kernel $\exp(-sh)$ to define the heat kernel regularization. 
Then define the expectation $\lra{\lra{\cdot}}$ by the formula  
\refp{7.1} with this new $h(\la,ta)$. As $q_1=q_1^g$, therefore we have
\be
\lra{\lra{d_{q_1(\la, ta)} X}} =0\;.
\ee{10.8}
However it may not be true that $q_2 = q^g_2$, so it may not be true that
$\lra{\lra{d_{q_2(\la, ta)} X}}$ vanishes.  As before, with $a^2=I$, define
\be
\fj (\la,a,g) = \lra{\lra{Ja}}\; . \ee{10.9}
In this case, we establish as in the proof of Theorem VI that
\be
\fj (\la,a,g) = \fz^{Q_j(\la)} (a,g)\; . \ee{10.10}
Thus the proof of Theorem I shows:

\mbni{Theorem VII.} {\it Let $a\in\A$, assume $a^2=I$, and also assume 
the regularity hypotheses on $Q_j(\la)$ and $d_1 a=[Q_1 (\la),a]$, stated 
in \S XI.  Then the expectation $\fz^{Q_j(\la)}(a,g)$ is 
independent of $\la$.}

\setcounter{equation}{0}
\section{Regularity Hypotheses}
As explained in the introduction, our results depend crucially on some 
regularity hypotheses.  In order for $\fz$ to exist, we assume 
$e^{-H(\la)}=e^{-Q(\la)^2}$ exists and is trace class on $\CH$.
We give sufficient conditions to ensure this, as well as to ensure the
validity of the results claimed in \S I --- IX.  The content of \S X 
require only minor modification of these hypotheses.  We have explored
the consequences of these hypotheses in [QHA].

\begin {itemize}
\ritem{1.}
The operator $Q$ is self-adjoint operator on $\CH$, odd with respect to
$\gamma$, and $e^{-\beta Q^2}$ is trace class for all $\beta>0$.

\ritem{2.} 
For $\la\in\Lambda$, where $\Lambda$ is a open interval on the real line,
the operator $Q(\la)$ can be expressed as a perturbation of $Q$ in the form
\be
Q(\la)=Q+W(\la)\;.
\ee{11.2}
Each $W(\la)$ is a symmetric operator on the domain $\D=C^\infty(Q)$.

\ritem{3.} Let $\la$ lie in any compact subinterval $\Lambda'\subset\Lambda$. 
The inequality 
\be
W(\la)^2 \le a Q^2 + bI\;,
\ee{11.3}
holds as an inequality for forms on $\D\times\D$.  The
constants $a<1$ and $b<\infty$ are independent of $\la$ in the compact 
set $\Lambda'\subset\Lambda$.

\ritem{4.} Let $R=(Q^2 + I)^{-1/2}$. The operator $Z(\la)=R W(\la) R$ 
is bounded uniformly for $\la\in\Lambda'$, and the difference quotient
\be
\frac{Z(\la)-Z(\la')}{\la-\la'}
\ee{11.4}
converges in norm to a limit as $\la'\to\la \in \Lambda'\subset\Lambda$.

\ritem{5.} The bilinear form $d_\la a$ satisfies the bound
\be
\Vert R^\alpha d_\la a R^\beta\Vert < M\;,
\ee{11.5}
with a constant $M$ independent of $\la$ for $\la\in\Lambda'$. 
Here $\alpha,\beta$ are non-negative constants and $\alpha+\beta <1$.
\end{itemize}

In certain examples we are interested in the behavior of $\fj(\la,a,g)$ 
as $\la$ tends to the boundary of $\Lambda$.  In this case, we 
may establish the constancy of $\fj$ with estimates that are weaker than
(1--5) at the endpoint of $\Lambda$, by directly proving the existence 
and continuity of $\fj$ at the endpoint.  We study one such an example
in [J], though other types of endpoint singularities are also of interest
(often involving a $\la\to \infty$ limit).

\end{document}